\documentclass[prl,twocolumn,lengthcheck,superscriptaddress,letterpaper,nofootinbib,nopreprintnumbers,showpacs]{revtex4-1}

\usepackage{color}
\usepackage{graphicx}
\usepackage{float}
\usepackage{amsmath}
\usepackage{acronym}
\usepackage{multirow}
\usepackage{lineno}
\usepackage{svn-multi}
\usepackage{import}
\usepackage{hyperref}
\usepackage[normalem]{ulem}

\newcommand{\msun}{\ensuremath{\mathrm{M}_\odot}}

\newcommand{\TheEvent}{GW150914}
\newcommand{\ground}{``Ground''}
\newcommand{\lisa}{``eLISA+Ground''}

\newcommand{\SPINCONE}{{\ensuremath{0.32_{-0.29}^{+0.49}}}} 
\newcommand{\SPINCTWO}{{\ensuremath{0.44_{-0.40}^{+0.50}}}} 
\newcommand{\MONESCOMPACT}{{\ensuremath{36_{-4}^{+5}}}} 
\newcommand{\DISTANCECOMPACT}{{\ensuremath{410_{-180}^{+160}}}} 
\newcommand{\MTWOSCOMPACT}{{\ensuremath{29_{-4}^{+4}}}} 

\begin{document}
\title{Multi-band gravitational-wave astronomy: parameter estimation and tests of general relativity with space and ground-based detectors}
\pacs{%
04.80.Nn, 
95.55.Ym,
04.25.dg, 
95.85.Sz, 
97.80.--d  
}
\author{Salvatore Vitale}
\affiliation{LIGO, Massachusetts Institute of Technology, Cambridge, Massachusetts 02139, USA}

\begin{abstract}
With the discovery of the black hole binary (BBH) coalescence GW150914 the era of gravitational-wave (GW) astronomy has started. 
It has recently been shown that BBH with masses comparable to or higher than GW150914 would be visible in the eLISA band a few years before they finally merge in the band of ground-based detectors. This would allow for pre-merger electromagnetic alerts, dramatically increasing the chances of a joint detection, if BBH are indeed luminous in the electromagnetic band. In this paper we explore a quite different aspect of multi-band GW astronomy, and verify if, and to what extent, measurement of masses and sky position with eLISA could improve parameter estimation and tests of general relativity with ground-based detectors. We generate a catalog of 200 BBH and find that having prior information from eLISA can reduce the uncertainty in the measurement of source distance and primary black hole spin by up to factor of 2 in ground-based GW detectors. The component masses estimate from eLISA will not be refined by the ground based detectors, whereas joint analysis will yield precise characterization of the newly formed black hole and improve consistency tests of general relativity.
\end{abstract}
\maketitle

\section{introduction}

The discovery of the gravitational wave (GW) event \TheEvent{}~\cite{GW150914-DETECTION} detected by the LIGO~\cite{TheLIGOScientific:2014jea} and Virgo~\cite{TheVirgo:2014hva} collaboration, has opened the era of gravitational-wave astronomy. \TheEvent{} was generated by a binary black hole (BBH) coalescence at {\DISTANCECOMPACT} Mpc . The masses of the two black holes have been estimated~\cite{GW150914-PARAMESTIM} to be {\MONESCOMPACT} \msun{} and {\MTWOSCOMPACT} \msun.
Only weak constraints have been set on the spins magnitude~\cite{GW150914-PARAMESTIM}, and little could be said about their orientation, a piece of information which could have helped pinpoint the formation channel of \TheEvent~\cite{GW150914-ASTRO,Vitale:2015}.
In fact, the 90\% confidence intervals (CI) on the dimensionless spin magnitudes spanned most of the prior support, with medians and 90\% CI given by {\SPINCONE} and {\SPINCTWO}~\cite{GW150914-PARAMESTIM}.

In a recent paper~\cite{Sesana} it has been underlined how events with masses like \TheEvent{} or higher would be visible in the eLISA band up to redshifts of $\sim0.4$, years before they finally coalesce in the band of ground-based detectors. Detections of orbital eccentricity with eLISA could help in distinguishing between black hole populations~\cite{BreivikPrep,2016arXiv160501341N}, and joint detections could improve existing bounds on dipole emission~\cite{2016arXiv160304075B}.

Ref.~\cite{Sesana} points out how joint detections would provide valuable information on the sky position of the source, with sky localization errors of only a few square degrees. The small size of the error areas, combined with the fact that electromagnetic (EM) facilities would know in advance the time of the event, would increase the chance of successfully finding the (potential) EM counterpart. Pre-merger alerts would make it possible to look for both pre and post-merger counterparts, if either is produced by a BBH~\cite{2016arXiv160203920C,2016arXiv160400955T,2016ApJ...819L..21L}.
In this paper, we look at a quite different aspect of multi-band GW astronomy: using the information from the eLISA analysis to inform the ground-based analysis. 
One might expect that if masses are already partially constrained by the eLISA observation of the early inspiral, the ground-based parameter estimation analysis could be more precise than in a blind approach. In this paper we quantify this improvement.

\section{Method}

The main goal of this paper is to compare the estimation of the physical parameters of heavy-BH binaries in two scenarios. We will use {\ground} to refer to results obtained only using ground-based detectors. In this case nothing is known about the sources. Conversely, we will use {\lisa} to refer to results obtained when the chirp mass and mass ratio, as well as the sky position, are somehow constrained by the eLISA observations \emph{before} analyzing the ground-based detectors data.

In practice eLISA might be able to put some weak constraints on spins and other parameters too. However we will be conservative and assume that only mass and sky position information will come from eLISA for those multi-band detections (work is ongoing to assess the measurability of heavy-mass BBH spins with eLISA~\cite{KleinPrep}).

We have generated a catalog of thousands of BBH binaries with masses such that they could be observed by eLISA. In particular, we uniformly generated component masses (in the source frame) in the range $[25-100]$\msun, with $M_{tot}\leq100$. This mass range is narrower than what is considered in~\cite{Sesana}. The main reason behind our choice, in particular for excluding lower mass events, is computational (analysis of lower mass events is more expensive). While~\cite{Sesana} only uses an analytic Fisher matrix approach, we cannot ignore computational cost while performing full Monte Carlo numerical simulations. We will discuss, later in the text, how our conclusions could be affected by this choice.

The dimensionless spin magnitude of each BH was uniformly drawn from the range $[0-0.99]$ (this is the range where the waveform family we used has been calibrated against numerical relativity~\cite{Hannam:2013oca,Khan:2015jqa}), while the spin tilt angles (i.e. the angle between the spin vector and the orbital angular momentum) were uniform in the unit sphere. The luminosity distances were random in comoving volume, using a $\Lambda$CDM flat cosmology~\cite{Ade:2015xua}.
Since eLISA will be online after the end of this decade, we worked with a plausible ground-based detector network for the 2020's, that is: two LIGOs in the US~\cite{TheLIGOScientific:2014jea,0264-9381-27-8-084006}, Virgo in Italy~\cite{TheVirgo:2014hva}, one LIGO in India~\cite{M1100296} and KAGRA in Japan~\cite{PhysRevD.88.043007}. However, since we will be comparing parameter estimation accuracies with and without the eLISA information, our primary result is largely insensitive to exact details of the future ground-based detector network.
For the masses and network configuration we considered, the distribution of sources producing a network signal-to-noise ratio (SNR) of 10 or more in the ground-based network would peak at a distance of $\sim 3.5$ Gpc ($z\simeq 0.6$). However, since sources farther than $z\simeq 0.4$ would not be detectable with eLISA~\cite{Sesana}, from the catalog of events generated as explained above we only kept sources with redshift smaller than 0.4.

200 BBH sources are drawn from the restricted set, and used in this paper. 

We performed parameter estimation with the IMRPhenomPv2 waveform approximant~\cite{Hannam:2013oca,Khan:2015jqa} that was used to estimate the parameters of {\TheEvent}. We worked with the nested sampling flavor of \texttt{lalinference}~\cite{LALInference}.  The algorithm we ran is thus identical to what used in~\cite{GW150914-PARAMESTIM} with a main difference: instead of sampling in the luminosity distance, we sampled directly using the redshift, which was assigned a prior uniform in comoving volume in the range $z\in [10^{-5}, 2]$. Note that due to the cosmological distances of these sources, the masses in the detector frame will be redshifted to higher values (by a factor $(1+z)$). Given our redshift range, redshifted masses in the detector frame take values in the range {$[25,180]$}\msun.
We did not marginalize over calibration errors, implicitly assuming that by the time eLISA is online the calibration of ground-based detectors will be better than one percent (current practical limits using the photon calibrator~\cite{2009CQGra..26x5011G} are $\sim0.8\%$~\cite{2016arXiv160203845T}).

The parameters of the signals were estimated first assuming no prior eLISA information. For those runs we used flat priors in the component masses in the range $[10, 250]$\msun, flat priors in the spin magnitude in the range $[0,0.99]$, uniform on the sphere for the orbit orientation, sky position, and spin orientation. These are the {\ground} results.

We then performed a second parameter estimation analysis (on the same signals) restricting the priors of masses and sky positions around their true values, assuming that eLISA will give the correct estimates for those parameters, within its error bars. For each event, we centered the prior of the chirp mass at the true value, with a range given by $\pm0.001$\% of the true value, the symmetric mass ratio with a range of $\pm 3\%$\footnote{Although we used the symmetric mass ratio $\eta$ in \texttt{lalinference}, in what follows we will report the asymmetric mass ratio $q\equiv m_2/m_1\leq1$.} and right ascension and declination with a range of $\pm 3^\circ$. Those numbers come from the most conservative values given in Fig.~3 of~\cite{Sesana}. These are the {\lisa} results.

Finally, to ensure that our findings would not be affected by unusual noise fluctuations, we worked with zero-noise realization~\cite{RodriguezEtAl:2013}. This consists of assuming that the noise is zero for each frequency bin, while still considering a colored advanced LIGO and Virgo power spectral density to calculate the likelihood~\cite{LALInference}. It has been shown that the results found with this approach are reliable, with corrections of the order of $1/\rho^3$, $\rho$ being the signal-to-noise ratio~\cite{PhysRevD.77.042001}.

\section{Results}

\begin{figure}[htb]
\centering
\includegraphics[width=0.98\columnwidth]{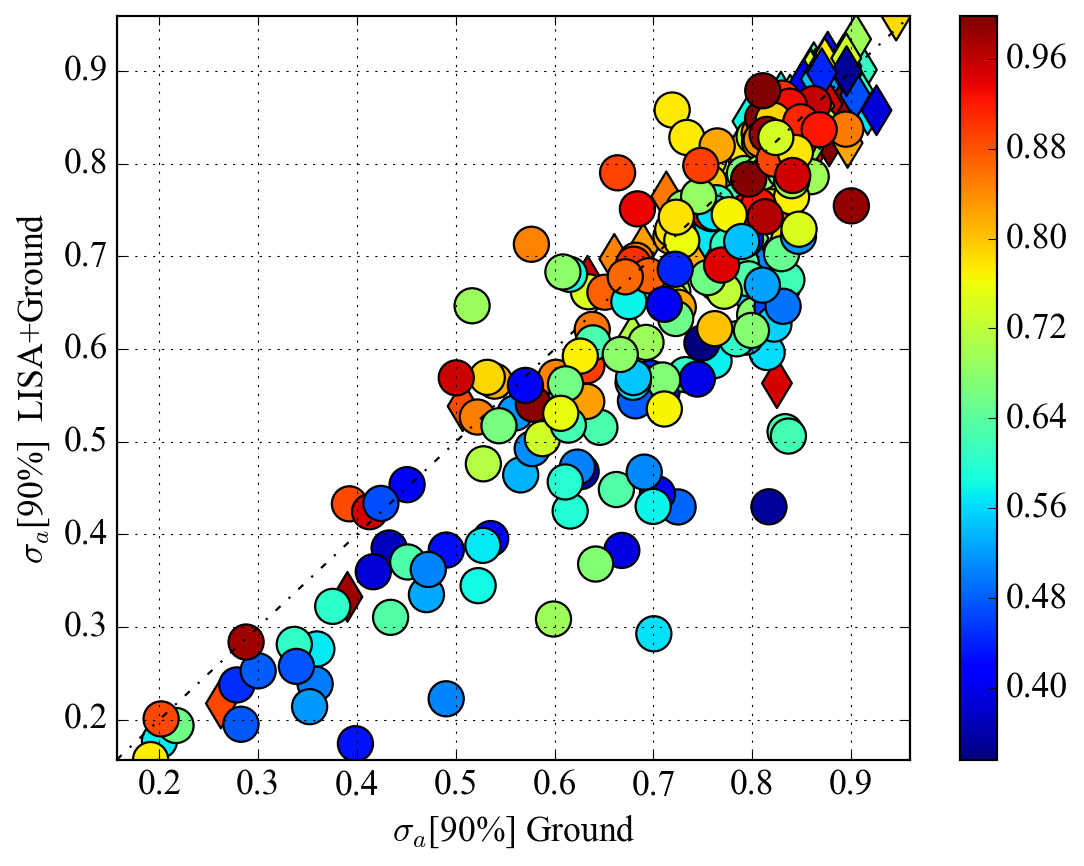}
\caption{90\% CI for the measurement of the spin magnitude for the primary (circles) and secondary BH (diamonds, mostly hidden underneath circles in the top right). The X axis reports the uncertainty only using ground-based detectors, while the Y axis uses prior eLISA mass and sky position estimates. The colorbar is the mass ratio (in the range [0,1]). It is clear how a join \lisa{} analysis can yield smaller uncertainties.}
 \label{Fig.1}
\end{figure}

In this section, we use 90\% CI to quote uncertainties. We will use the word ``primary'' and the index 1 for the most massive BH in the binary.
We look first at the measurement of the spin magnitudes. In Fig.~\ref{Fig.1} we show the uncertainty on the measurement of the primary BH spin magnitude $a_1$ (circles) and secondary BH spin  magnitude $a_2$ (diamonds) in the {\ground} analysis (X axis) and in the {\lisa} analysis (Y axis). The color bar reports the asymmetric mass ratio ($q\equiv m_2/m_1\leq1$).

The dashed line is the locus of points with equal uncertainties in both analyses. For most events, the {\lisa} analysis yields narrower posterior distributions. However scatter exists around the dashed line, especially for poorly estimated primary spins and most secondary spins. We notice that most points below the diagonal are blue, i.e. systems with large asymmetry in the masses (low mass ratio $q$) benefit more from having eLISA information. The reasons why some points are above the diagonal will be given later this section.

\begin{figure}[htb]
\centering
\includegraphics[width=\linewidth]{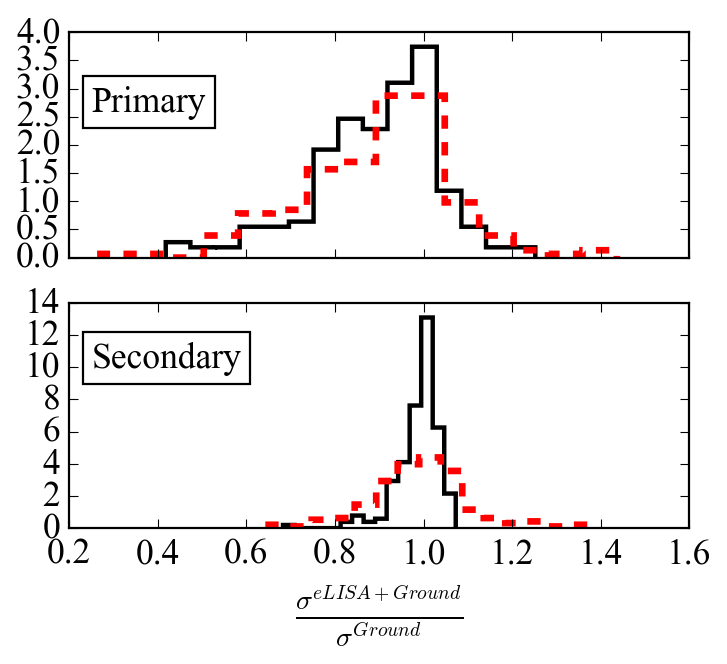}
\caption{Normalized distribution of the ratios between the 90\% CI in the {\lisa} and {\ground} analyses for the spin magnitude (full) and tilt angle (dashed). The top panel is for the primary (i.e. most massive) black hole, the bottom panel for the secondary.}
 \label{Fig.2}
\end{figure}

In Fig.~\ref{Fig.2} we report the ratio of the 90\% CI for the spin magnitude in the {\lisa} over the {\ground} analysis (continuous lines, top panel for primary BH, bottom for secondary BH). We see that for the primary BH in the best case the uncertainty is $\sim40\%$ of what would be obtained with a {\ground} analysis, while spin 2 uncertainties are generally unchanged (and large).

Measuring the orientation of spins in binaries could give important insights in the evolution of the systems, for example suggesting which formation channel (globular cluster or galactic field) is more common~\cite{GW150914-ASTRO,Vitale:2015}. In Fig.~\ref{Fig.2} we report the relative improvement in the \lisa analysis for the tilt angles (dashed lines) and find that on average the 90\% CI for the tilt angle of the primary (secondary) BH in the {\lisa} runs is $0.91$ ($0.99$) of the corresponding uncertainty in the {\ground} runs. For the tilt angle too, it is the case that more can be gained for the most massive (primary) black-hole in the binary.
One might be surprised that the distributions have some support above 1, i.e. that there are systems for which the joint {\lisa} analysis does \emph{worse} than the \ground. 

\begin{figure}[htb]
\centering
\includegraphics[width=\linewidth]{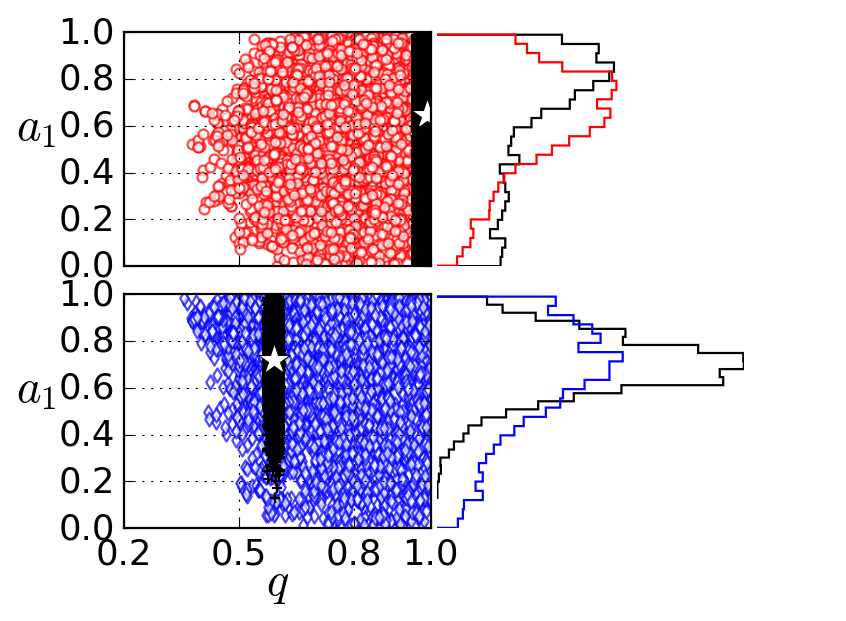}
\caption{(Left panels) 2D $a_1-q$ posterior distributions of two representative events. The colored points are the sample from the \ground{} runs, the black points from the \lisa{} analysis. The white star is the true value. (Right panels) Histograms of the corresponding marginalized $a_1$ distributions . When the true $q$ is much smaller than 1 (bottom row) the \lisa{} analysis does significantly better.}
 \label{Fig.3}
\end{figure}

To show why that happens one should look at the full, multidimensional, posterior distribution. In order to make things easier to visualize on paper, we shall focus on the joint 2D distribution of spin 1 magnitude ($a_1$) and mass ratio ($q$). This is shown for two representative events of our catalog in Fig~\ref{Fig.3}. In the left panels we show the 2D distribution of the {\ground} analysis (colored markers) and the corresponding {\lisa} runs (black markers, too dense to be resolved individually). The white stars are at the true value of the parameters. The histograms on the right show the marginalized 1D distribution for the primary spin magnitude (colored lines for \ground, black lines for \lisa). 

Due to correlation between the parameters, the $a_1-q$ posterior in the {\ground} analysis typically spans the whole range of spins for large mass ratios, while at lower $q$'s fewer values of the spins are supported. 

For the event on the top panel, the true value of $q$ is close to one, i.e on a region of the 2D parameter space where most values of $a_1$ are supported. 
Since the \lisa{} run only explores that side of the parameter space (black markers), the resulting $a_1$ marginalized distribution will be quite broad (top-right black histogram). 
For the same event, the \ground{} analysis (red markers) will also explore the low-$q$ region, where spins are mostly in the middle of the prior range. This will make the $a_1$ posterior to look narrower in the \ground{} analysis (top-right red histogram).

The opposite happens if the mass ratio is small (bottom panel). The {\lisa} analysis will be cutting a narrow rectangle in an area of the 2D plot where fewer spins are allowed, whereas the full {\ground} run (blue markers) will explore the  high-$q$ region, which will broaden the $a_1$ distribution (bottom histogram). 

When it comes to the component masses, ground-based cannot add much, while they can benefit from the eLISA information to get e.g. better estimates of the final BH mass and spins, see below.  For the 200 BBH we considered, the relative uncertainty (defined as the ratio of the 90\% CI over the true value) of the \ground{} analysis would span the range $\sim[10-50]\%$ for $m_1$ and $\sim[15-120]\%$ for $m_2$ \footnote{We note that the relative uncertainties for $m_2$ are larger since the true values are smaller, by convention.}, whereas for eLISA alone they are a few percent at most~\cite{Sesana}.
These are the typical uncertainties we can expect for heavy BBH binaries from ground-based detectors \emph{at any time} (i.e. not a function of the number of detectors). For example GW150914 , with an SNR of $\sim25$, had 90\% CI uncertainties on the component masses of $\sim25\%$~\cite{GW150914-PARAMESTIM}. Joint \lisa{} detections would thus prove extremely useful in inferring the mass function on heavy black-holes.

Even though the estimation of the component spins might not always be better in the {\lisa} analysis, the uncertainty on the \emph{final} BH's spin does always get reduced, since that also depends on the component masses uncertainties which are drastically smaller in \lisa. We have used the same numerical relativity methods~\cite{Healy:2014yta} used in~\cite{GW150914-PARAMESTIM} to calculate the final spin, and found that on average the 90\% CI for the final spin in the {\lisa} analysis are $43\%$ of the corresponding intervals for the {\ground} estimates. For the same reasons, the estimate of the final BH mass in the \lisa{} analysis is significantly better, on average 4 times smaller. 

Even for the 5-detector ground-based network we considered, \lisa{} will not improve upon the sky localization from eLISA alone (eLISA will benefit from having modulation in the signal while orbiting for months or years). This statement might change if more ground-based detectors are online. Finally, estimation of luminosity distance is improved in the \lisa{} analysis, with 90\% CI which are on average $87\%$ (and often $60-70\%$) of the \ground{} runs. This might help for cosmology, if EM counterparts are indeed found.

In general relativity (GR), the phase of the GW waveform can be constructed as a power series in the frequency whose coefficients are determined either by the post-Newtonian (PN) theory or by calibration over numerical simulations (e.g. \cite{KhanEtAl:2016}). We have complemented our study by performing consistency tests of general relativity on 25 random systems, considering the test coefficients used in~\cite{GW150914-TESTOFGR} (Tab. I).

These tests are performed by allowing these, otherwise fixed, coefficients to vary around their GR value, and measure them, together with the ``usual'' GR parameters.
For all the test parameters, the uncertainties are significantly narrower in the \lisa{} analysis. Significant improvements are obtained for the early-inspiral test parameters ($\delta\hat\varphi_0$ (0~PN) to $\delta\hat\varphi_3$ (1.5~PN)) which benefit the most from the chirp mass being precisely estimated from the eLISA analysis. For some events we observed factors of 5 improvement, although the average improvement is smaller ($\sim85\%$). However, eLISA itself might already  measure low PN parameters, since it has access to the very early-inspiral phase. It is more interesting to check if late-stage inspiral and merger-ringdown parameters ($\delta\hat\beta$'s and $\delta\hat\alpha$'s) are improved in the joint analysis, since these parameters are measured from the last few cycles, not accessible to eLISA. We observed a small average improvement of $\sim98\%$ for the $\delta\hat\alpha$'s and $\sim92\%$ for the $\delta\hat\beta$'s. For all test parameters we observed that the improvement is larger for asymmetric mass systems. Joint \lisa{} signals will thus be extremely valuable both for tests on individual events and for tests across multiple sources~\cite{LiEtAl:2012a,Agathos:2013upa}. 
Furthermore, with a joint \lisa{} analysis we will measure with extreme precision sky position, mass and spin of the newly formed BH, which could help search for axion clouds around BHs~\cite{2016arXiv160403958A}. Given that eLISA will measure the initial masses very precisely, while ground-based detectors would measure the last few cycles (and hence final mass and spins directly from the ringdown) we expect that inspiral-merger-ringdown consistency tests~\cite{GW150914-TESTOFGR,Ghosh:2015xx} will also dramatically benefit from joint events. We leave the quantification of these improvements for a future publication.
    
\section{Conclusions}

In has been suggested that heavy binary black holes, such as GW150914 will enable multi-band gravitational-wave astronomy, since they will be visible in both space and ground-based detectors. In this article we consider a catalog of such BBH and compare the precision in estimating their parameters in two scenarios: a) using the data from ground-based detectors alone (the two LIGOs, Virgo, LIGO India and KAGRA) and b) using some information from the earlier eLISA analysis to restrict the priors on the ground-based detector analysis. Following~\cite{Sesana}, we have made the conservative assumption that eLISA would provide estimates of the chirp mass (with a $10^{-3}\%$ uncertainty), mass ratio ($3\%$) and sky position ($3^\circ$ for both rightascension and declination). We have found that systems with larger mass ratios can benefit more from previous eLISA information. For those events, the 90\% confidence interval estimates on spins and distance could get up to a factor of {2} better. For systems with mass ratios close to unity, the benefit will be smaller. The mass estimates from eLISA for BBH will be much more precise than what the ground-based detectors will achieve (few percent versus few tens of percent). This implies that the properties of the newly formed BH will be known much better for \lisa{} events.
We have performed consistency tests of general relativity~\cite{GW150914-TESTOFGR} on a subset of 25 BBH, and found that all test parameters benefit from eLISA mass estimates, with asymmetric systems improving more. We stress that we have been quite conservative in our choice of the uncertainties from eLISA, and that the actual benefit might be even larger than what we find here.
In conclusion, events with joint eLISA and ground-based estimation have the potential to boost our understanding of heavy black holes, their formation channels, and general relativity.
 
\section{Acknowledgments}

The author would like to thank Walter Del Pozzo, Matt Evans, Erik Katsavounidis, Natalia Korsakova, Georgia Mansell, Rai Weiss and the CBC group of LIGO for useful suggestions. The author acknowledges useful discussions with E.~Barausse, E.~Berti, K.~Breivik, V.~Kalogera and N.~Yunes.
Thanks to the anonymous PRL referee for pointing out that the redshift distribution was too broad in the first draft of this work.
The author acknowledges the support of the National Science Foundation and the LIGO Laboratory.
LIGO was constructed by the California Institute of Technology and Massachusetts Institute of Technology with funding from the National Science Foundation and operates under cooperative agreement PHY-0757058.
The author would like to acknowledge the LIGO Data Grid clusters, without which the simulations could not have been performed. 
Specifically, we thank the Albert Einstein Institute in Hannover, supported by the Max-Planck-Gesellschaft, for use of the Atlas high-performance computing cluster. A special thanks to the system administrator of the Atlas Cluster, Carsten Aulbert.

This is LIGO document number P1600138.

\bibliography{cbc-group,GW150914_refs,draft}
\end{document}